\documentclass[aps,prb,amsmath,superscriptaddress,a4paper,floatfix,showpacs,twocolumn]{revtex4}
\usepackage{graphicx}
\usepackage{amsmath}
\usepackage{amsfonts}
\usepackage{amssymb}
\usepackage{xcolor}

\begin{document}

\title{Breakdown of the escape dynamics in Josephson junctions}

\author{D. Massarotti}
\email{dmassarotti@na.infn.it}
\affiliation{Universit\'{a} degli Studi di Napoli "Federico II", Dipartimento di Fisica, via Cinthia, 80126 Napoli, Italy}
\affiliation{CNR-SPIN UOS Napoli, Complesso Universitario di Monte Sant'Angelo, via Cinthia, 80126 Napoli, Italy}
\author{D. Stornaiuolo}
\affiliation{Universit\'{a} degli Studi di Napoli "Federico II", Dipartimento di Fisica, via Cinthia, 80126 Napoli, Italy}
\affiliation{CNR-SPIN UOS Napoli, Complesso Universitario di Monte Sant'Angelo, via Cinthia, 80126 Napoli, Italy}

\author{P. Lucignano}
\affiliation{CNR-SPIN UOS Napoli, Complesso Universitario di Monte Sant'Angelo, via Cinthia, 80126 Napoli, Italy}

\author{L. Galletti}
\affiliation{Universit\'{a} degli Studi di Napoli "Federico II", Dipartimento di Fisica, via Cinthia, 80126 Napoli, Italy}
\affiliation{CNR-SPIN UOS Napoli, Complesso Universitario di Monte Sant'Angelo, via Cinthia, 80126 Napoli, Italy}

\author{D. Born}
\affiliation{Leibniz Institute of Photonic Technology e.V., P.O. Box 100239, D-07702 Jena, Germany}

\author{G. Rotoli}
\affiliation{Seconda Universit\'{a} di Napoli, Dipartimento di Ingegneria Industriale e dell'Informazione, via Roma 29, 81031 Aversa (Ce), Italy}

\author{F. Lombardi}
\affiliation{Department of Microtechnology and Nanoscience, MC2, Chalmers University of Technology, S-41296 G\"{o}teborg, Sweden}

\author{L. Longobardi}
\affiliation{Seconda Universit\'{a} di Napoli, Dipartimento di Ingegneria Industriale e dell'Informazione, via Roma 29, 81031 Aversa (Ce), Italy}
\affiliation{American Physical Society, 1 Research Road, Ridge, New York 11961, USA}

\author{A. Tagliacozzo}
\affiliation{Universit\'{a} degli Studi di Napoli "Federico II", Dipartimento di Fisica, via Cinthia, 80126 Napoli, Italy}
\affiliation{CNR-SPIN UOS Napoli, Complesso Universitario di Monte Sant'Angelo, via Cinthia, 80126 Napoli, Italy}

\author{F. Tafuri}
\affiliation{CNR-SPIN UOS Napoli, Complesso Universitario di Monte Sant'Angelo, via Cinthia, 80126 Napoli, Italy}
\affiliation{Seconda Universit\'{a} di Napoli, Dipartimento di Ingegneria Industriale e dell'Informazione, via Roma 29, 81031 Aversa (Ce), Italy}

\date{\today}

\begin{abstract}

We have identified anomalous behavior of the escape rate out of the zero-voltage state in Josephson junctions with a high critical current density $J_c$. For this study we have employed YBa$_2$Cu$_3$O$_{7-x}$ grain boundary junctions, which span a wide range of $J_c$ and have appropriate electro-dynamical parameters. Such high $J_c$ junctions, when hysteretic, do not switch from the superconducting to the normal state following the expected stochastic Josephson distribution, despite having standard Josephson properties such as a Fraunhofer magnetic field pattern. The switching current distributions (SCDs) are consistent with non-equilibrium dynamics taking place on a local rather than a global scale. This means that macroscopic quantum phenomena seem to be practically unattainable for high $J_c$ junctions. We argue that SCDs are an accurate means to measure non-equilibrium effects.  This transition from global to local dynamics is of relevance for all kinds of weak links, including the emergent family of nano-hybrid Josephson junctions. Therefore caution should be applied in the use of such junctions in, for instance, the search for Majorana fermions. 

\end{abstract}

\pacs{74.50.+r, 85.25.Cp, 74.40.Gh, 74.78.Na}

\maketitle

\section{Introduction}
Rich physics, such as the Josephson effect\cite{josephson}, quantum coherence\cite{likharev,leggettrmp} and quantum interference\cite{jaklevic}, arise when two coherent quantum systems are weakly coupled. Some examples of systems that exhibit such physics are superconducting Josephson junctions\cite{barone}, the flow of superfluid $^4$He through an array of nano-apertures\cite{hokinson}, and the observation of the Josephson effect in Bose-Einstein condensates\cite{levy}. Superconductors have traditionally held a special place as test-bench systems for these kinds of quantum phenomena due to the ease of scaling and integrating them into real quantum devices, offering a high degree of measurability and control of a macroscopic wave function.

Every experiment or application using a superconducting weak link is based on how the phase difference $\varphi$ between the electrodes evolves in time and space\cite{josephson,barone,likharev,libroBez}. The large variety of barriers now available between the superconducting electrodes offer novel functionalities and efficient tuning of physical processes occurring at the nanoscale and at different interfaces\cite{defranceschi2,doh,topological,morpurgo,bouchiat}. A recent example is the proposal to use the Josephson effect for the detection of the Majorana fermions\cite{kane}. 

\begin{figure*}[htbp]
\begin{center}
\includegraphics[width=0.9\linewidth]{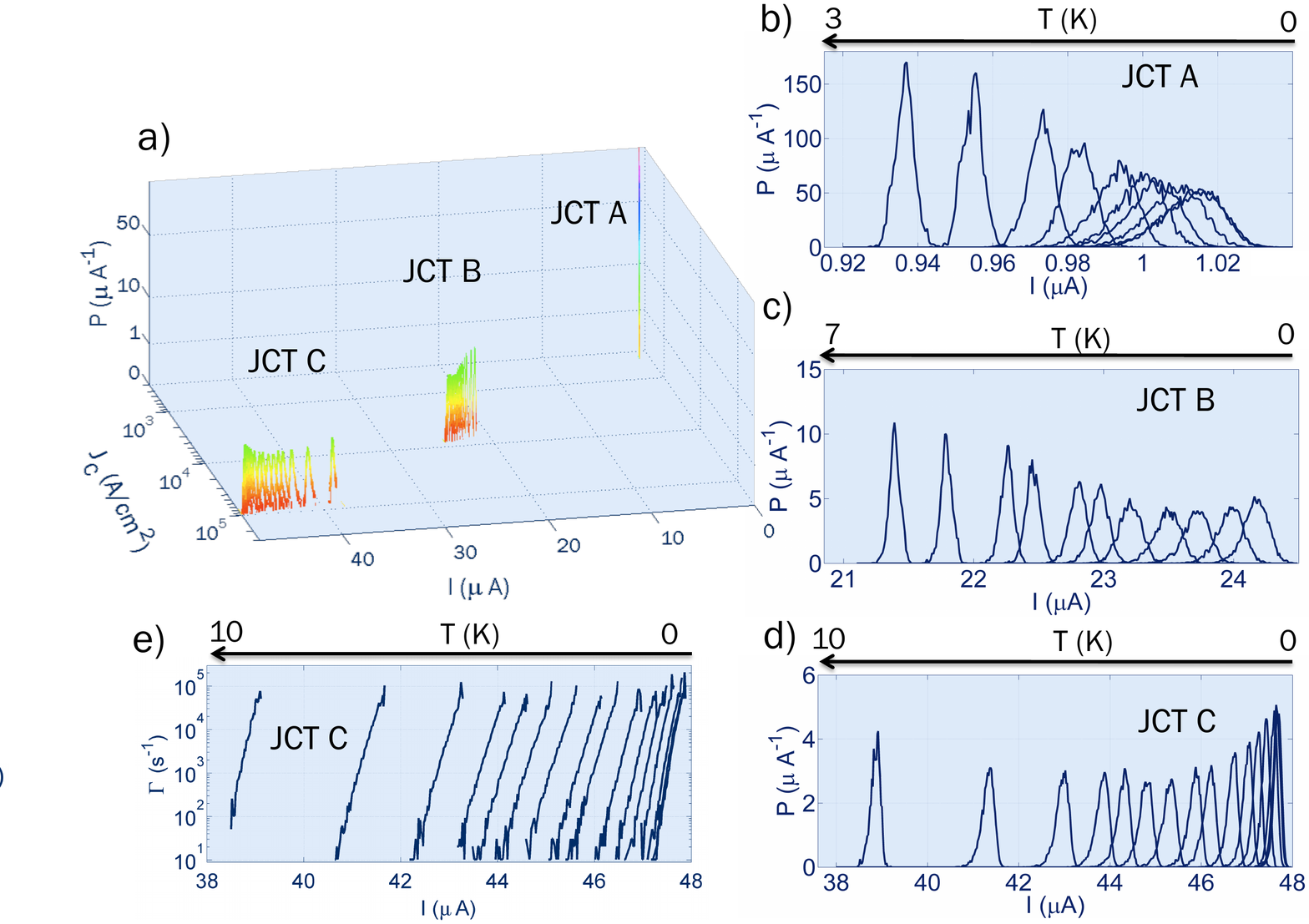}
\caption{a) Three-dimensional representation of the SCDs measured for various temperatures on three GB YBCO JJs. b) At low $J_c$ values (5$\cdot$10$^2$ A/cm$^2$), the SCDs of JCT A are confined to a small range of currents, with $\sigma$ of the order of 10 nA. With increasing $J_c$ the histograms progressively cover a larger interval of currents, both absolutely and relatively (JCTs B and C in panels c and d respectively). In panel e) the escape rate curves $\Gamma(I)$ computed from the SCDs of JCT C have been shown for the same temperature range as in panel d.}
\label{figure1}
\end{center}
\end{figure*}

Progress in material science in producing a larger variety of interfaces and in nanotechnologies applied to superconductivity, is promoting a rethinking of the phase dynamics of Josephson junctions (JJs). Here we give evidence of a breakdown of a fundamental tenet of the Josephson effect: the transition  from the superconducting to the normal state does not follow the expected stochastic Josephson phase dynamics, but has a more intriguing balance between local and global energy processes. The onset of non-equilibrium effects is the key to describe the local processes, which may radically change perspectives on how to interpret the physics and make predictions on the performances of a large variety of "smart" Josephson devices. We measure switching current distributions (SCDs) which codify the very general process of the escape of a particle (phase) from a potential well in a JJ\cite{barone,likharev}, keeping track in our case of non-equilibrium effects. Roughly speaking, SCDs are obtained in JJs with hysteretic current-voltage (I-V) characteristics by counting the number of times the system switches from the superconducting to the resistive state within a small window of bias current, when ramping forth and back the bias current. Thermally activated processes are well understood in JJs both in the underdamped\cite{kurk,fulton,jackel} and in the moderately damped\cite{kautz_secondo} regime. The transition to the macroscopic quantum tunneling regime has been theoretically\cite{Leggett,Leggett_b,Leggett_c} and experimentally\cite{webb2,devoret1985,bauch_2005} widely investigated. SCD measurements focus on "the very moment" at which resistance originates in superconducting weak links. Thus we use the power of encoding information of non-equilibrium local processes in fluctuations to characterize these dynamical processes.

In the resistively and capacitively shunted junction (RCSJ) model\cite{barone}, the damping parameter $Q=\omega_p RC$ is proportional to the square root of $I_c$ via the plasma frequency $\omega_p=(2eI_c/\hbar C)^{1/2}$ at zero bias current, where $R$ and $C$ are the resistance and capacitance, respectively. In a more general approach, $Q$ has a frequency dependence\cite{kautz_secondo,dani_PRB}, which includes the effects of the external shunting impedance. A junction cannot sustain an unlimited increase in the critical current $I_c$ and thus in the quality factor $Q$ through larger critical current density $J_c$ while still preserving all the properties of the Josephson effect and all the features of the underdamped regime in the I-V  curves.

The breakdown of the Josephson dynamics is found at higher values of the critical current density. The classical Josephson phase dynamics, which takes place in junctions characterized by lower critical current densities $J_c$, are replaced at high $J_c$ values by a regime driven by non-equilibrium dynamics where phase information is lost. Non-equilibrium effects produce hysteretic I-V characteristics and modify the influence of dissipation, thus becoming measurable through modeling of the SCD in terms of heating modes. The heating modes follow a heat diffusion-type equation in analogy to phase slip events\cite{phase_slips,langer}. In the present work, this transition from classical to non-equilibrium phase dynamics is found for high critical temperature superconductors (HTS) grain boundary (GB) junctions, but should be expected for any kind of JJs\cite{kleinsasser1,kleinsasser2}. Specific thresholds may depend on the type of junctions and materials, but the features of the transition are universal. HTS GB junctions are the ideal system to identify this transition because of the possibility of varying the critical supercurrent density $J_c$ over a wide range of values\cite{hans,rop}.

\begin{figure*}[htbp]
\begin{center}
\includegraphics[width=0.9\linewidth]{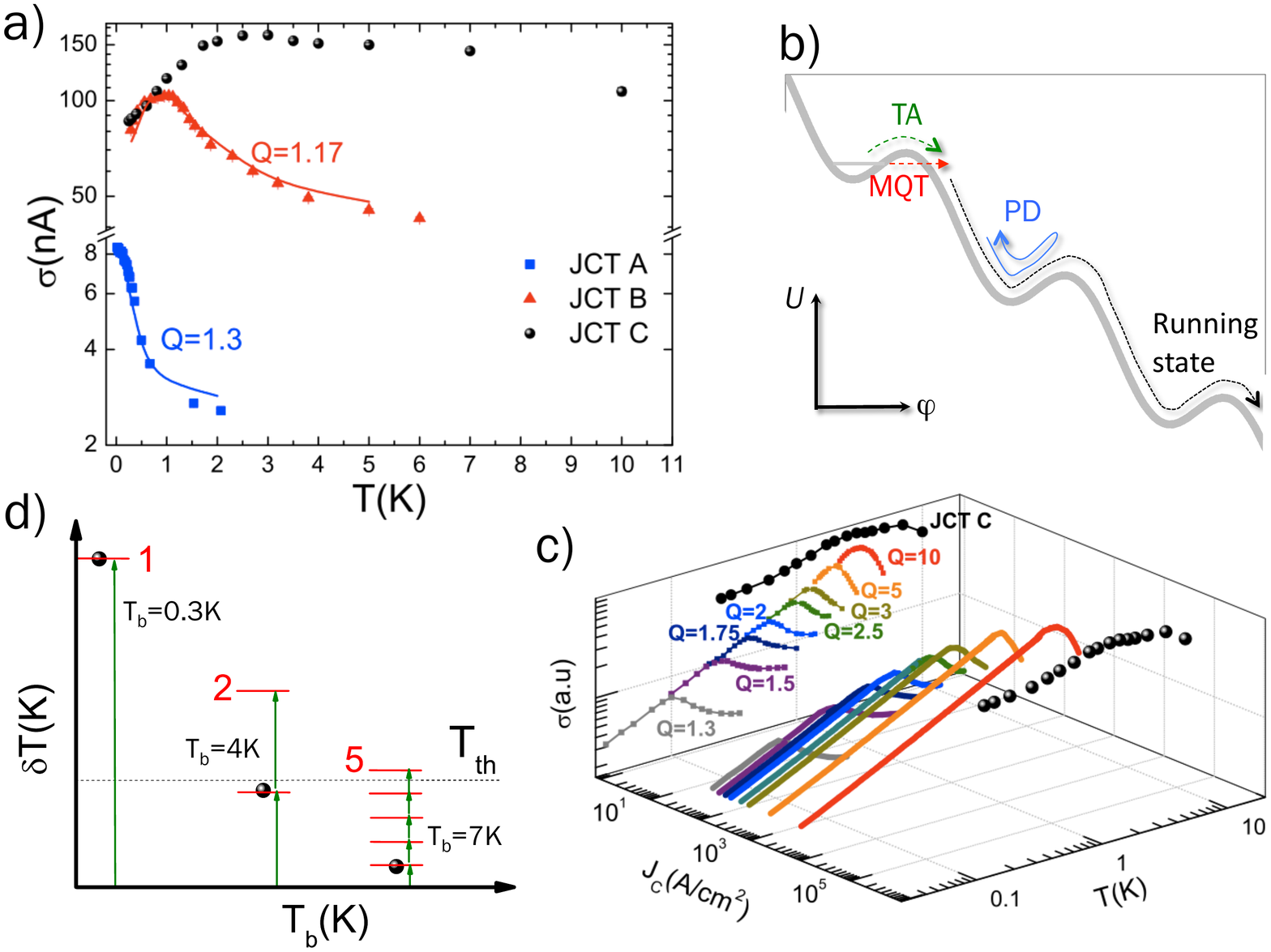}
\caption{a) Temperature dependence of $\sigma$ measured on GB JCTs A (blue squares), B (red triangles) and C (black points) respectively. The blue and red solid lines are Monte Carlo simulations of the phase dynamics, according to multiple escape and retrapping processes in the washboard potential (shown in panel b), with $Q$ = 1.30 and 1.17 for JCTs A and B respectively. Some data from sample A have been presented in Ref. \onlinecite{luigi2} previously. c) Simulations of the thermal dependence of $\sigma$ as a function of $J_c$ for different values of the $Q$ damping parameter (full color lines), confined to the moderately damped regime, are compared with experimental data of JCT C (black points). Keeping all the other junction parameters fixed, an enhancement of $J_c$ leads to an increase of $I_c$ and $Q$. An increase in $Q$ produces steeper $\sigma$ tails above $T^*$ and cannot reproduce the broadened experimental data of JCT C. This is even more evident in the two dimensional ($\sigma - T$) projection. For completeness, in the ($\sigma - T$) projection the color points referred to the results of Monte Carlo simulations have been reported for $Q$ values ranging from 1 to 10, and compared with the data of JCT C (black points). The colored lines are guides for the eye. At high $J_c$  the Langevin approach does not hold anymore and non-equilibrium concepts apply. Simulations fitting the SCDs of JCT C (see Fig. \ref{figure4}) are consistent with multiple heating events as reported pictorially in panel d. As explained in the text, $T_b$ is the bath temperature and $T_{th}$ is the threshold temperature above which the transition to the resistive state occurs. In panel d) the arrows qualitatively sketch the temperature jump due to a single heating event and the number of events necessary to induce the switching to the resistive state.}
\label{figure2}
\end{center}
\end{figure*}

\section{Josephson phase dynamics}

YBa$_2$Cu$_3$O$_{7-x}$ (YBCO) off axis biepitaxial GB junctions provide a large variety of transport regimes because of d-wave order parameter effects as well as the modulation of the barrier transparency through different relative orientation of the electrodes\cite{hans,nuovo,rop}. Details on the preparation and properties of the GB junctions used in this work are given in Refs. \cite{rop,supplemental,LSAT}. We present here data from junctions (JCT) with low (JCT A), intermediate (JCT B) and high (JCT C) values of $J_c$. For each junction $J_c$ has been determined as the ratio between the measured $I_c$ in the I-V curves and the geometrical area of the junction (width $\times$ thickness of the YBCO film)\cite{supplemental}. For JCT C we have employed nanofabrication techniques to be in the appropriate range of $I_c$\cite{dani_PRB,gustafsson2}. To study their escape rates we have thermally anchored the samples to the mixing chamber of a He$^3$/He$^4$ Oxford dilution refrigerator and performed measurements of SCDs. Measurement procedures and filtering used in this experiment are described elsewhere\cite{luigi1}.

Switching histograms of JCTs A, B and C are shown in Fig. \ref{figure1}. The 3-D view gives an intuitive picture: SCDs cover distinct current ranges, and when $J_c$ increases, they become broader. In the right part of the figure, each set of SCDs is displayed with appropriate scales for a better view of the details of their temperature dependence. In Fig. \ref{figure1}e the escape rate $\Gamma$ out of the zero-voltage state as a function of the bias current $I$ has been plotted for JCT C in the same temperature range. $\Gamma(I)$ curves have been computed from the SCDs following Fulton and Dunkleberger \cite{fulton}. The standard deviation $\sigma$ of the histograms is reported for all junctions in Fig. \ref{figure2}a. The combined analysis of $\sigma$, of the skewness $\gamma=m_3 /\sigma^3$ ($m_3$ being the third central moment of the distribution), and of their temperature evolution characterize the phase dynamics. 

The histograms from JCTs A and B match well the predictions of the RCSJ model for moderately damped JJs. Low values of $I_c$\cite{kautz_secondo,kivioja2005,mannik,luigi1,vion}, especially in HTS d-wave junctions, where intrinsic sources contribute to dissipation\cite{dani_PRB,luigi2}, lead to the moderately damped (MD) regime ($Q\simeq1$). In the MD regime the SCDs become narrower with increasing temperature for temperatures above $T^*$, which is defined as the temperature at which $\sigma(T)$ is largest (in the case of JCT B $T^*$ is about 1.0 K). This is often reported in the literature as the phase diffusion (PD) regime\cite{kautz_secondo,kivioja2005,mannik,luigi1,luigi2}, indicating a phase diffusive dynamics due to multiple escape and retrapping events in the washboard potential (sketched in Fig. \ref{figure2}b), or equivalently described in terms of thermal activation above a dissipation barrier\cite{vion}. Below $T^*$, SCDs obey the thermal activation (TA) regime. Upon further lowering the temperature the junction can eventually enter the macroscopic quantum tunneling regime\cite{devoret1985,luigi2,Leggett,Leggett_b,Leggett_c} (see Fig. \ref{figure1}b, JCT A). The MD regime gives a very important reference for our discussion of sample C. In both cases the width of the SCD decreases upon increasing the temperature, as shown in Figs. \ref{figure1} and \ref{figure2}. Only an exact fitting of the SCDs at different temperatures can reveal the very different physics occurring in the various junctions.

JCTs A and B of the present work have been chosen since they are fabricated on different substrates (see Ref. \onlinecite{supplemental}) and both show the transition to the PD regime, in which the derivative $d\sigma/dT$ is negative, but with a different detailed $T$ dependence\cite{luigi1}. The choice of the substrate affects the effective capacitance of the circuit the junction is embedded into\cite{pekolaLTP}. As a consequence of this choice, for JCT B we observed the transition from TA to the PD regime, while for JCT A the TA regime is completely suppressed and we observed a direct transition from MQT to the PD regime\cite{luigi2}. We show below that the “anomalous” thermal behavior of high $J_c$ JCT C cannot be explained in terms of frequency dependent damping or as a consequence of the shell circuit, and the numerical simulations of Fig. \ref{figure2}c fully support this.

\begin{figure}
\begin{center}
\includegraphics[width=\linewidth]{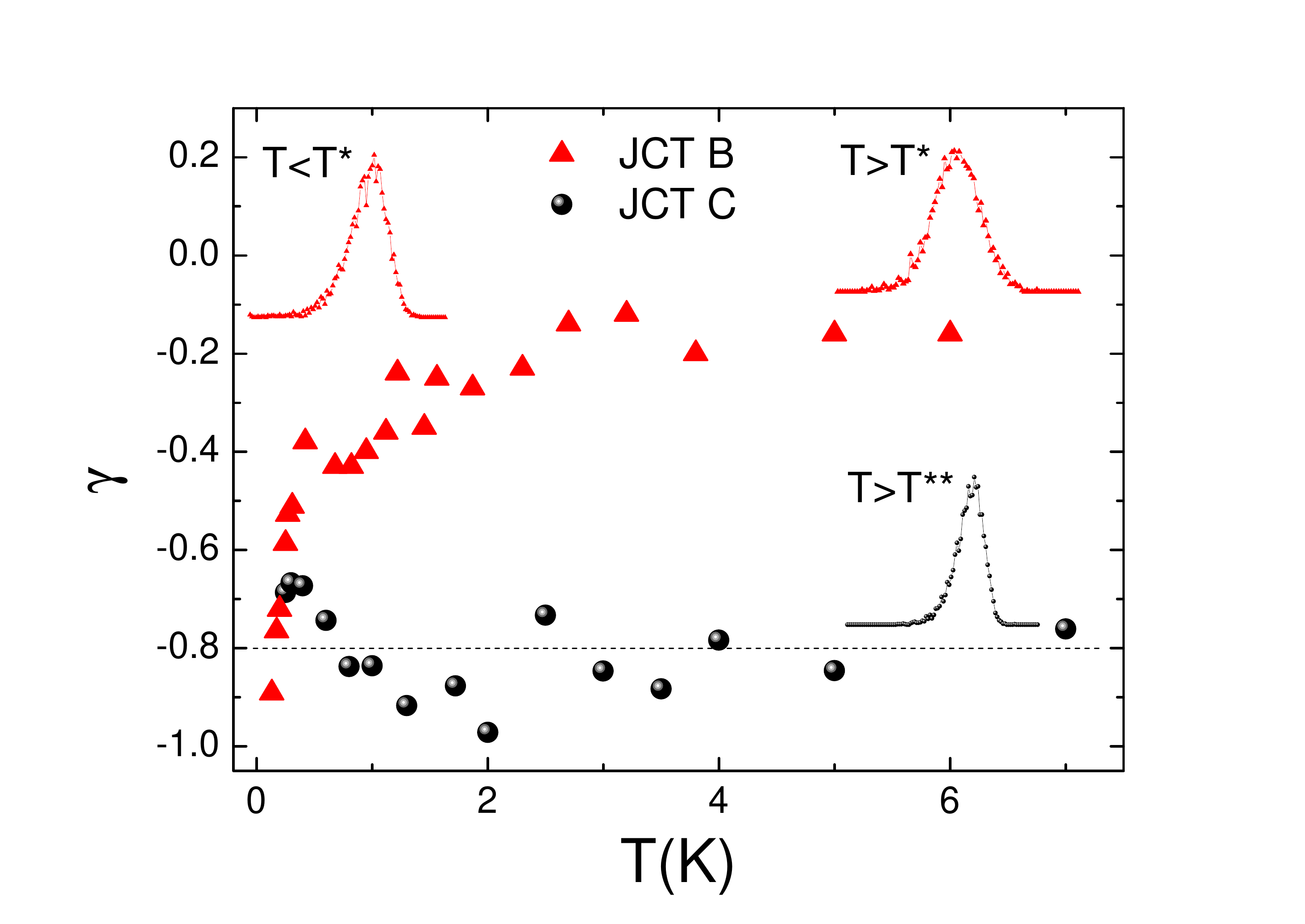}
\caption{Temperature dependence of the skewness $\gamma$ of the SCDs measured on JCTs B (red triangles) and C (black circles) respectively. Retrapping processes in the PD regime cause a progressive symmetrization of the SCDs of JCT B, signaled by the thermal dependence of $\gamma$, which is almost 0 above the transition temperature $T^*$. Asymmetric SCDs, as the histogram measured at $T=0.1K<T^*$ (shown in the top left corner of the figure), become Gaussian-like distributions above $T^*$ (see the histogram measured at $T=4K>T^*$ in the top right corner). The histograms of JCT C are asymmetric over the whole temperature range, including for temperatures above $T^{**}$ (as shown by the SCD measured at $T=7$ K reported in the lower right corner of the figure). The black dashed line indicates the mean value of the skewness ($\gamma \simeq -0.8$).}
\label{figure3}
\end{center}
\end{figure}

In JCTs A and B the rate of decrease in $\sigma$ above $T^*$ with increasing $T$ is well described by the Monte Carlo fit of the phase dynamics\cite{supplemental} (blue and red solid lines in Fig. \ref{figure2}a, respectively). The temperature dependence of the skewness $\gamma$ gives additional distinctive criteria to confirm the intermediate dissipation regime\cite{luigi1,Bezryadin_ultimo}. This is signaled by a transition for JCT B for instance from $\gamma$ $\simeq$ -0.9 at low temperatures ($T \simeq$ 0.1 K) in the TA to $\gamma$ $\simeq$ -0.1 above the transition temperature $T^* \simeq$ 1.0 K in the PD, indicating a progressive symmetrization of the SCDs\cite{luigi1} (see Fig. \ref{figure3}). Therefore, the moderately damped regime gives the opportunity of comparing $\sigma(T)$ and $\gamma(T)$ dependencies of SCD spectra at higher temperatures, which turn out as unambiguous distinctive criteria for the different switching modes.

JCT C is characterized by high values of $J_c$  close to those observed in nanowires\cite{sahu,chang}. This device exhibits radically different phase dynamics above 3 K, which represents a transition temperature we indicate as $T^{**}$. Also in the case of JCT C, $T^{**}$ is defined as the maximum of $\sigma(T)$, but we prefer to use a different symbol to stress the distinct switching dynamics, as pointed out below. The rate of decrease of $\sigma$ above $T^{**}$ turns to be a distinctive marker of the phase dynamics. In device C the slope of $\sigma(T)$ above $T^{**}$ is much smaller when compared to those of JCTs A\cite{luigi2} and B and of all moderately damped JJs. The smooth decrease of $\sigma$ for $T > T^{**}$ cannot be described in terms of the intermediate dissipation regime, as evident from the numerical simulations reported in Fig. \ref{figure2}c. According to the RCSJ model, keeping all the other junctions parameters fixed, an increase of $J_c$ leads to an enhancement of $I_c$ and $Q$. The increase of $Q$ moves $T^*$ to higher values, and the negative slope of $\sigma(T)$ above $T^*$ becomes steeper and steeper (as shown by the numerical simulations for $Q$ ranging between 1 and 10 reported in Fig. \ref{figure2}c).

In addition, while moderately damped JJs show a progressive symmetrization of the switching histograms near to and above $T^*$ (see JCT B in Fig. \ref{figure3}), SCDs of JCT C are asymmetric over the entire temperature range. In JCT C, the $\gamma$ factor is temperature independent, consistent with what is observed in pure phase slip systems\cite{Bezryadin_ultimo} (see Fig. \ref{figure3}, the black dashed line indicates the mean value $\gamma \simeq$-0.8 for JCT C). These behaviors are quite distinct and do not fall in the framework of any regime of the RCSJ model. 

\section{Switching dynamics of high critical current density Josephson junctions}

\begin{figure}[htbp]
\begin{center}
\includegraphics[width=0.8\linewidth]{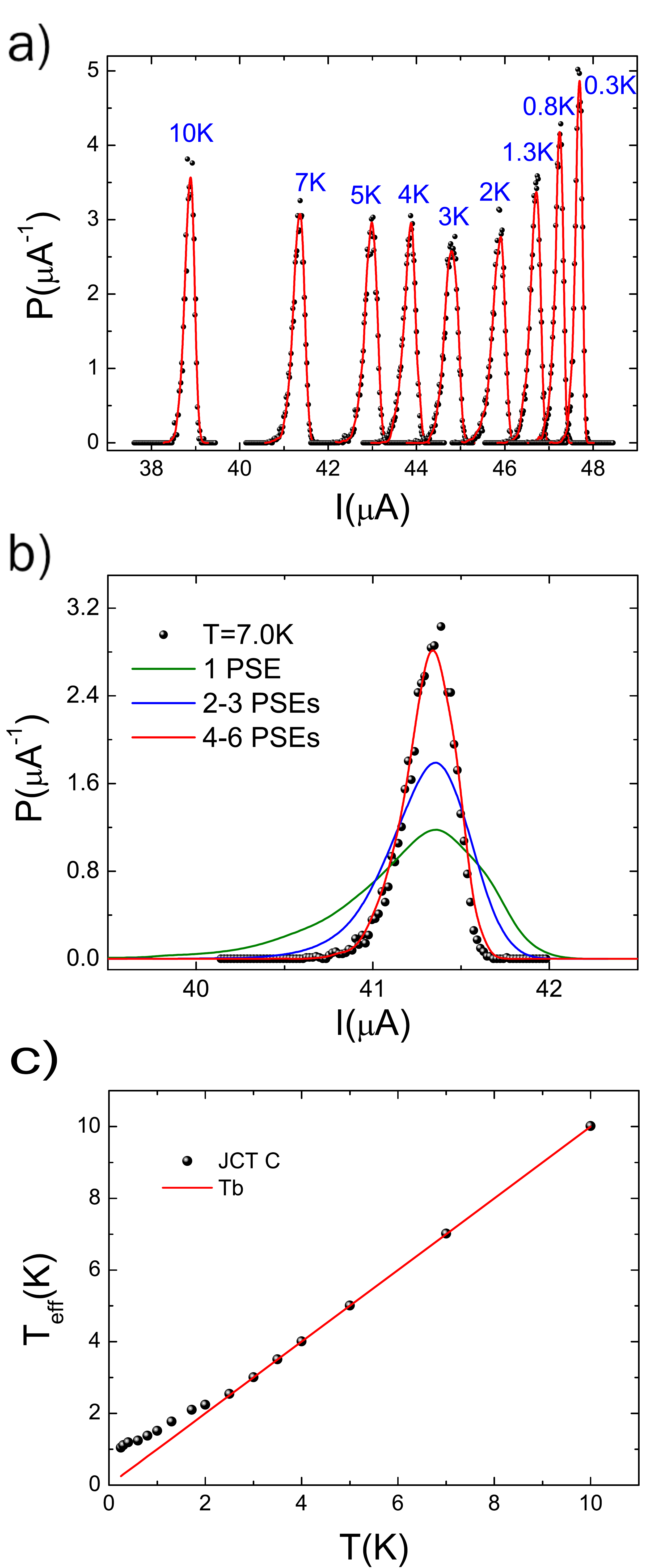}
\caption{a) SCDs measured on JCT C (black circles) along with the fit (red solid lines) resulting from numerical simulations of heat diffusion-like dynamics, according to eq. \ref{diffeq}. b) Fitting procedure above $T^{**}$: single heating event dynamics does not reproduce the experimental histogram, signaling a multiple PSE process. Dissipation due to heat relaxation plays a relevant role. c) Temperature dependence of $T_{eff}$ estimated from numerical simulations of eq. 1. Below $T^{**}$, a single heating event induces the switching and $T_{eff} > T_b$. Above $T^{**}$, $T_{eff}$ and $T_b$ coincide since the system is able to thermalize in the time interval between well separated heating events, and multiple PSEs are needed to overcome $T_{th}$.}
\label{figure4}
\end{center}
\end{figure}

We find that the numerical simulation of a transition driven by local heating events accounts well for devices in the $J_c$ interval (10$^4$ A/cm$^2$-10$^5$ A/cm$^2$), as shown for JCT C in Fig. \ref{figure4}a. Hysteresis in I-V curves\cite{doh,pekola2,jung} does not necessarily indicate canonical Josephson phase dynamics, even in the presence of a Fraunhofer magnetic field pattern\cite{kleinsasser2}. It may rather arise as a result of local heating processes, possibly induced by intrinsic inhomogeneous composition unavoidable for high $J_c$ junctions. Large values of Joule power density deposited in the weak link can induce a self-heating process during the switch to the resistive branch\cite{pekola2}. The absence of a set of self-consistent electrodynamics parameters to describe JCT C is a strong indication of the failure of the standard Josephson dynamics. This failure is of general relevance, applying both to conventional low $T_c$ JJs\cite{kleinsasser1,kleinsasser2} and to the emergent class of hybrid nanoscale junctions\cite{doh,morpurgo,bouchiat}. For larger values of $J_c$, heating driven mechanisms become dominant with a transition to the normal state locally in the junction area. These events are the mirror of non-equilibrium processes and can be modeled as "phase slips entities" (PSEs), as confirmed by the details of the simulations which are reported in Refs. \onlinecite{supplemental,goldbart,lau,Leggett_RMP}, in the sense that they are local processes, break the coherence of the phase information and are described by a heat diffusion-like equation. In particular, the probability for a single heating event can be still described in terms of the Langer-Ambegaokar-McCumber-Halperin (LAMH) theory\cite{langer} and further extensions\cite{zaikin}, and PSEs can still be approximately assumed to be far apart in time.

The typical time scale of a PSE is of the order of the Ginzburg Landau relaxation time $\sim \tau (u) = ( 1-u)^{-1} \: \tau _0 $ where $\tau _0 = \pi \hbar /8 k_B T_c  $ and $u=
T/T_c <1$, with $T_c$  the superconducting critical temperature. The energy dissipated by a PSE $E_{PSE} (I) =  \phi _o I $ is rather high, as the modulus of the order parameter has to vanish over a length $ \xi (T)= \xi_0 (1-u^4)^{1/2}/( 1-u^2)$, where  $\xi _0 $ is the zero temperature coherence length in the superconductor and $\phi _o$ is the flux quantum. Following an approach proposed in Ref. \onlinecite{goldbart} for low critical temperature superconducting (LTS) wires, our numerical simulation of the temperature jump induced by a PSE obeys the phenomenological diffusive equation for the relaxation of the temperature gradient:

\begin {equation}
\frac{d\delta T}{dt} + \alpha \left ( T,T_b \right ) ~ \delta T  = r(T_b, t )+  \eta (T,I) \sum_{i}\delta(t-t_i)    \: .
\label{diffeq}
\end {equation}

Here  $\delta T = T-T_b $ is the deviation from the temperature of the bath $T_b$. The relaxation coefficient $\alpha \left (T,T_b \right )$ depends on the thermal conductivity $K(T)$, on the thermal capacity ${\cal { C}}_v(T)$ of the phase slip volume, and on $T_b$\cite{supplemental}. $ r(T_b, t )$ is the noise source due to the environment with an admittance $Y( \omega ) $, while  $\eta(T,I )$  is the temperature jump due to the PSEs which occur at the stocasticaly distributed times $t_i$\cite{supplemental}. $\eta(T,I )$ is implicitly defined by the following equation:

 \begin{equation}
   \phi _o  \: I  = \int ^{ T +{ \eta} (T,I ) }_{T}  \!\!\!\!\! \!\!\!\!\! \!\!\!\!\! dT' \:\:  {\cal { C}}_v(T') .
     \label{jump}
 \end{equation}

In Fig. \ref{figure4}a we show the SCDs derived from the experiment on JCT C over a wide temperature range, $T \in [0.25K,10K]$. The continuous red curves correspond to the fit obtained integrating eq. \ref{diffeq}. To quantify how many heating events are needed to escape to the finite voltage state, we define the threshold temperature $T_{th}$, i.e. the temperature above which the system is definitely out of the zero voltage state. At low temperatures, $T< T^{**}$, our fitting procedure does not depend on $T_{th}$. This is the regime when a single heating event is enough to drive the transition. Following Ref. \onlinecite{goldbart}, we assume that both the specific heat and the thermal conductivity $K(T)$ are weighted averages of their BCS and Fermi liquid limits. Two main effects discriminate between the low temperature and the high temperature behavior.

The temperature jump $\eta(T)$ depends on temperature because the specific heat is strongly temperature dependent. At low temperatures the specific heat is quite low, thus with each PSE there is a considerable increase in the temperature. In addition, the thermal conductivity and the thermal capacity are both quite low as the system is deeply into the superconducting phase.  In a rough sense, the system is rather isolated from the environment and the thermal shock due to a heating event is destructive for the superconducting state. For this reason a single heating event can induce a direct jump to the resistive state at low temperatures: it induces a large local heating that is difficult to dissipate. The system is not at equilibrium with its environment, and we can define an effective temperature $T_{eff}$ for the system, which is higher than $T_b$ as shown in Fig. \ref{figure4}c.

At high temperatures we are in the opposite regime of small $\eta$(T) per heating event\cite{supplemental}. In addition both the thermal conductivity $K(T)$ and the thermal capacity ${\cal { C}}_v(T)$ increase with increasing temperature as well. Thermal diffusion and contact with the environment is more effective and multiple PSEs are required for switching. This occurs above $T^{**}$, where the derivative $d\sigma/dT$ is negative. $T_{eff}$ and $T_b$ coincide above $T^{**}$ (see Fig. \ref{figure4}c), which we interpret as the temperature at which the system is able to thermalize during the time interval between well separated heating events. In this temperature range, the value chosen for the fitting parameter $T_{th}$ becomes important to quantify the number of successive PSEs, which are responsible for the transition. In Fig. \ref{figure4}b the difference between a single and multiple heating events is shown. At 7 K for instance about 4-6 PSEs are needed to reproduce the experimental SCD. Experimental SCDs complemented by the numerical simulations follow the passage from single to multiple heating events. This is similar to what has been observed in LTS  nanowires\cite{sahu,chang,goldbart}. A consistent set of the junction parameters (temperature jump $\eta$, number of heating events) can be extracted from these simulations, as discussed in Ref. \onlinecite{supplemental}.

\section{Discussion and concluding remarks}

The passage above a characteristic $J_c$ threshold to a state that does not sustain Josephson phase-coherence should be universally expected in weak link systems.  The threshold probably depends on the type of junction, responding to structurally different ways of enhancing barrier transparency and generating sources of local heating. The threshold of $10^4-10^5$ A/cm$^2$ found for our junctions is consistent with the general empiric rule on HTS GBs\cite{hans,rop}. In bicrystal junctions, an increase in the misorientation angle between the two electrodes determines a well-known decrease of the critical current density. For small misorientation angles devices are found in the flux-flow regime. For $\theta > 20^\circ$ Josephson phenomenology starts to appear for $J_c$ values of $10^3$ A/cm$^2$ \cite{hans,rop}. Several effects have also been observed in conventional trilayer Nb/Al-Oxide/Nb JJs above the same threshold of about $10^4$ A/cm$^2$ that could be attributed to non-equilibrium dynamics\cite{kleinsasser1,kleinsasser2}.

Some analogies can be established with what is observed in $^4$He superfluid\cite{hokinson}. Here the passage from weak to strong coupling manifests itself in a change in the current-phase relation [$I(\varphi)$]. In the strong coupling regime, where the healing length of the superfluid is lower than the diameter of the constriction, deviations from the $\sin \varphi$ relation appear, while sinusoidal Josephson oscillations have been measured in the opposite limit\cite{hokinson}. An increase in $J_c$ and in the coupling between the electrodes leads to the presence of other harmonics in the $I(\varphi)$, which might become multivalued\cite{likharev}.

A heat diffusion-like model breaking phase-coherent information is consistent with our data on high $J_c$ JCT C. Here the switch to the normal state is accompanied by a local release of energy characteristic of a PSE. When departing from the supercurrent branch, non-equilibrium processes produce an unexpected heating. This is surprisingly different from what commonly is accepted for hysteretic Josephson junctions, where heating can only influence the retrapping phenomena, as a memory of the history of heating in the resistive state\cite{Leggett_RMP,castellano}.

In the framework of the RCSJ model the resistance arises from non-local properties well described by the (frequency dependent) quality factor $Q$ of the circuit. This reminds us of the Landauer picture of quantum transport, in which quantum interference acts at the interface while the dissipation and memory loss occurs in the contacts. On the other hand, when the switch is driven by PSEs, it is the local properties of the order parameter which are important: phase memory is lost when the modulus of the order parameter becomes zero. The model based on local heating events can be in principle further extended to extract the $I(\varphi)$ and to fully define the role of dissipation in high $J_c$ junctions. 

To conclude, standard phase dynamics of a hysteretic Josephson junction collapses and cannot be sustained above some threshold of the critical current density $J_c$. In high $J_c$ devices information is lost in non-equilibrium dynamics, and can be partly codified in the local heating process. This is of great relevance for all the experiments using low-dimensional barriers, which should be concerned about possible heating effects, leading to distorted phase information. Non-equilibrium effects would obviously invalidate a large number of key predictions for Majorana fermions and nanoscale superconductivity in Josephson junctions, since for instance they can give rise to zero bias anomalies in conductance measurements\cite{argaman1}. The proof of quantifiable non-equilibrium processes already in the thermal regime poses severe constraints on the possible occurrence of macroscopic quantum phenomena at lower temperatures in high $J_c$ samples through standard SCD measurements.

\begin{acknowledgments}

The authors are indebted to Professor A. J. Leggett and Professor J. R. Kirtley for inspiring conversations and for a careful reading of the manuscript. The authors acknowledge financial support by Marie Curie Grant No. 248933, by Progetto FIRB HybridNanoDev RBFR1236VV, and by COST Action MP1201 (NanoSC COST).

\end{acknowledgments}


\begin{thebibliography}{40}

\bibitem{josephson} B. D. Josephson, Phys. Lett. \textbf{1}, 251 (1962).
    
\bibitem{likharev} K. K. Likharev, Rev. Mod. Phys. \textbf{51}, 101 (1979).

\bibitem{leggettrmp} A. J. Leggett, S. Chakravarty, A. T. Dorsey, Matthew P. A. Fisher, Anupam Garg, and W. Zwerger, Rev. Mod. Phys. \textbf{59}, 1 (1987).

\bibitem{jaklevic} R. C. Jaklevic, J. Lambe, A. H. Silver, and J. E. Mercereau, Phys. Rev. Lett. \textbf{12}, 159 (1964); Y. Sato and R. E. Packard, Rep. Prog. Phys. \textbf{75}, 016401 (2012).

\bibitem{barone} A. Barone and G. Patern{\`{o}}, \emph{Physics and Applications of the Josephson Effect} (John Wiley \& Sons, New York, 1982).

\bibitem{hokinson} E. Hoskinson, Y. Sato, I. Hahn, and R. E. Packard, Nat. Phys. \textbf{2}, 23 (2006).

\bibitem{levy} S. Levy, E. Lahoud, I. Shomroni, and J. Steinhauer, Nature \textbf{449}, 579 (2007); F. S. Cataliotti, S. Burger, C. Fort, P. Maddaloni, F. Minardi, A. Trombettoni, A. Smerzi, and M. Inguscio, Science \textbf{293}, 843 (2001).

\bibitem{libroBez} A. Bezryadin, \emph{Superconductivity in Nanowires} (Wiley, New York, 2012).

\bibitem{defranceschi2} S. De Franceschi, L. Kouwenhoven, C. Sch\"onenberger,	and W. Wernsdorfer, Nat. Nanotechnol. \textbf{5}, 703 (2010).

\bibitem{doh} Y. J. Doh, J. A. van Dam, A. L. Roest, E. P. A. M. Bakkers, L. P. Kouwenhoven, and S. De Franceschi, Science \textbf{309}, 272 (2005).

\bibitem{topological} M. Veldhorst, M. Snelder, M. Hoek, T. Gang, V. K. Guduru, X. L. Wang, U. Zeitler, W. G. v. d. Wiel, A. A. Golubov, H. Hilgenkamp, and A. Brinkman, Nat. Mater. \textbf{11}, 417 (2012); L. Galletti, S. Charpentier, M. Iavarone, P. Lucignano, D. Massarotti, R. Arpaia, Y. Suzuki, K. Kadowaki, T. Bauch, A. Tagliacozzo, F. Tafuri, and F. Lombardi, Phys. Rev. B \textbf{89}, 134512 (2014);  C. Kurter, A. D. K. Finck, P. Ghaemi, Y. S. Hor, and D. J. Van Harlingen, Phys. Rev. B \textbf{90}, 014501 (2014); I. Sochnikov, L. Maier, C. A. Watson, J. R. Kirtley, C. Gould, G. Tkachov, E. M. Hankiewicz, C. Br\"une, H. Buhmann, L. W. Molenkamp, and K. A. Moler, Phys. Rev. Lett. \textbf{114}, 066801 (2015).

\bibitem{morpurgo} H. B. Heersche, P. Jarillo-Herrero, J. B. Oostinga, L. M. K. Vandersypen, and A. F. Morpurgo, Nature \textbf{446}, 56 (2007).

\bibitem{bouchiat} J. P. Cleuziou, W. Wernsdorfer, V. Bouchiat, T. Ondarcuhu, and M. Monthioux, Nat. Nanotechnol. \textbf{1}, 53 (2006).

\bibitem{kane} L. Fu and C. L. Kane, Phys. Rev. Lett. \textbf{100}, 096407 (2008).

\bibitem{fulton} T. A. Fulton and L. N. Dunkleberger, Phys. Rev. B \textbf{9}, 4760 (1974).

\bibitem{kurk} J. Kurkij\"arvi, Phys. Rev. B \textbf{6}, 832 (1972).

\bibitem{jackel} L. D. Jackel, W. W. Webb, J. E. Lukens, and S. S. Pei, Phys. Rev. B \textbf{9}, 115 (1974).

\bibitem{kautz_secondo} R. L. Kautz and J. M.  Martinis, Phys. Rev. B \textbf{42}, 9903 (1990).

\bibitem{Leggett_b} A. O. Caldeira and A. J. Leggett, Phys. Rev. Lett. \textbf{46}, 211 (1981). 

\bibitem{Leggett} A. J. Leggett, J. Phys. Colloq. \textbf{39}, C6-1264 (1978).

\bibitem{Leggett_c} A. O. Caldeira and A. J. Leggett, Ann. Phys. \textbf{149}, 374 (1983).
 
\bibitem{webb2} R. F. Voss and R. A. Webb, Phys. Rev. Lett. \textbf{47}, 265 (1981); L. D. Jackel, J. P. Gordon, E. L. Hu, R. E. Howard, L. A. Fetter, D. M. Tennant, R. W. Epworth, and J. Kurkij\"arvi, Phys. Rev. Lett. \textbf{47}, 697 (1981); S. Washburn, R. A. Webb, R. F. Voss, and S. M. Faris, Phys. Rev. Lett. \textbf{54}, 2712 (1985).

\bibitem{devoret1985} J. M. Martinis, M. H. Devoret, and J. Clarke, Phys. Rev. B \textbf{35}, 4682 (1987).

\bibitem{bauch_2005} T. Bauch, F. Lombardi, F. Tafuri, A. Barone, G. Rotoli, P. Delsing, and T. Claeson, Phys. Rev. Lett. \textbf{94}, 087003 (2005); F. Lombardi, T. Bauch, J. Johansson, K. Cedergren, T. Lindstr\"om, F. Tafuri, and E. Stepantsov, Physica C \textbf{435}, 8 (2006).

\bibitem{dani_PRB} D. Stornaiuolo, G. Rotoli, D. Massarotti, F. Carillo, L. Longobardi, F. Beltram, and F. Tafuri, Phys. Rev. B \textbf{87}, 134517 (2013).

\bibitem{phase_slips} W. A. Little, Phys. Rev. \textbf{156}, 396 (1967).

\bibitem{langer} J. S. Langer and V. Ambegaokar, Phys. Rev. \textbf{164}, 498 (1967); D. E. McCumber and B. I. Halperin, Phys. Rev. B \textbf{1}, 1054 (1970)

\bibitem{kleinsasser1} A. W. Kleinsasser and R. A. Buhrman, Appl. Phys. Lett. \textbf{37}, 841 (1980).

\bibitem{kleinsasser2} R. E. Miller, W. H. Mallison, A. W. Kleinsasser, K. A. Delin, and E. M. Macedo, Appl. Phys. Lett. \textbf{63}, 10 (1993).

\bibitem{hans} H. Hilgenkamp and J. Mannhart, Rev.  Mod. Phys. \textbf{74}, 485 (2002).

\bibitem{rop} F. Tafuri and J. R. Kirtley, Rep. Prog. Phys. \textbf{68}, 2573 (2005).

\bibitem{nuovo} F. Lombardi, F. Tafuri, F. Ricci, F. Miletto Granozio, A. Barone, G. Testa, E . Sarnelli, J. R. Kirtley, and C. C. Tsuei, Phys. Rev. Lett. \textbf{89}, 207001 (2002).

\bibitem{supplemental} See Supplemental Material for the realization of grain boundary biepitaxial YBCO Josephson junctions and for numerical simulations of the escape dynamics.

\bibitem{LSAT} D. Stornaiuolo, G. Papari, N. Cennamo, F. Carillo, L. Longobardi, D. Massarotti, A. Barone, and F. Tafuri, Supercond. Sci. Technol. \textbf{24}, 045008 (2011).

\bibitem{gustafsson2} D. Gustafsson, D. Golubev, M. Fogelstr\"om, T. Claeson, S. Kubatkin, T. Bauch, and F. Lombardi, Nat. Nanotechnol. \textbf{8}, 25 (2013).

\bibitem{luigi1} L. Longobardi, D. Massarotti, G. Rotoli, D. Stornaiuolo, G. Papari, A. Kawakami, G.P. Pepe, A. Barone, and F. Tafuri, Phys. Rev. B \textbf{84}, 184504 (2011); D. Massarotti, L. Longobardi, L. Galletti, D. Stornaiuolo, D. Montemurro, G. Pepe, G. Rotoli, A. Barone, and F. Tafuri, Low Temp. Phys. \textbf{38}, 263 (2012).

\bibitem{kivioja2005} J. M. Kivioja, T. E. Nieminen, J. Claudon, O. Buisson, F. W. J. Hekking, and J. P. Pekola, Phys. Rev. Lett. \textbf{94}, 247002 (2005).

\bibitem{mannik} J. M\"annik, S. Li, W. Qiu, W. Chen, V. Patel, S. Han, and J. E. Lukens, Phys. Rev. B \textbf{71}, 220509 (2005).

\bibitem{vion} D. Vion, M. G\"otz, P. Joyez, D. Esteve, and M. H. Devoret, Phys. Rev. Lett. \textbf{77}, 3435 (1996).

\bibitem{luigi2} L. Longobardi, D. Massarotti, D. Stornaiuolo, L. Galletti, G. Rotoli, F. Lombardi, and F. Tafuri, Phys. Rev. Lett. \textbf{109}, 050601 (2012).

\bibitem{pekolaLTP} Y. Yoon, S. Gasparinetti, M. M\"ött\"onen, and J. P. Pekola, J. Low. Temp. Phys. \textbf{163}, 164 (2011).

\bibitem{Bezryadin_ultimo} A. Murphy, P. Weinberg, T. Aref, U. C. Coskun, V. Vakaryuk, A. Levchenko, and A. Bezryadin, Phys. Rev. Lett. \textbf{110}, 247001 (2013).

\bibitem{sahu} M. Sahu, M.-H. Bae, A. Rogachev, D. Pekker, T. C. Wei, N. Shah, P. M. Goldbart, and A. Bezryadin, Nat. Phys. \textbf{5}, 503 (2009).

\bibitem{chang} P. Li, P. M. Wu, Y. Bomze, I. V. Borzenets, G. Finkelstein, and A. M. Chang, Phys. Rev. Lett. \textbf{107}, 137004 (2011).

\bibitem{pekola2} H. Courtois, M. Meschke, J. T. Peltonen, and J. P. Pekola, Phys. Rev. Lett. \textbf{101}, 067002 (2008).

\bibitem{jung} M. Jung, H. Noh, Y. J. Doh, W. Song, Y. Chong, M. S. Choi, Y. Yoo, K. Seo, N. Kim, B. C. Woo, B. Kim, and J. Kim, ACS Nano \textbf{5}, 2271 (2011).

\bibitem{goldbart} N. Shah, D. Pekker, and P. M. Goldbart, Phys. Rev. Lett. \textbf{101}, 207001 (2008).

\bibitem{lau} M. Tinkham, J. U. Free, C. N. Lau, and N. Markovic, Phys. Rev. B \textbf{68}, 134515 (2003).

\bibitem{Leggett_RMP} Y. C. Chen, M. P. A. Fischer, and A. J. Leggett, J. Appl. Phys. \textbf{64}, 3119 (1988).

\bibitem{zaikin} D. S. Golubev and A. D. Zaikin, Phys. Rev. B \textbf{78}, 144502 (2008).

\bibitem{castellano} M. G. Castellano, G. Torrioli, F. Chiarello, C. Cosmelli, and P. Carelli, J. Appl. Phys. \textbf{86}, 6405 (1999).

\bibitem{argaman1} K. W. Lehnert, J. G. E. Harris, S. J. Allen, and N. Argaman, Superlattices Microstruct. \textbf{25}, 839 (1999); K. W. Lehnert, N. Argaman, H. R. Blank, K. C. Wong, S. J. Allen, E. L. Hu, and H. Kroemer, Phys. Rev. Lett. \textbf{82}, 1265 (1999).

\end{thebibliography}
\end{document}